\documentclass[prl,showpacs,amsmath,a4paper,twocolumn,superscriptaddress]{revtex4}
\usepackage{geometry}
\usepackage{amssymb}
\usepackage{amsmath}
\usepackage{graphicx}
\usepackage{amsfonts}
\usepackage{hyperref}
\usepackage{oldgerm}
\usepackage{color}
\usepackage{epsfig}
\usepackage{bbm}

\DeclareMathOperator{\Tr}{Tr}

\begin{document}
\title{The structure of multidimensional entanglement in multipartite systems}
\author{Marcus Huber}
\affiliation{University of Bristol, Department of Mathematics, Bristol, BS8 1TW, U.K.}
\affiliation{ICFO-Institut de Ciencies Fotoniques, 08860 Castelldefels, Barcelona, Spain}
\affiliation{Universitat Autonoma de Barcelona, 08193 Bellaterra, Barcelona, Spain}
\author{Julio I. de Vicente}
\affiliation{Institut f\"ur Theoretische Physik, Universit\"at Innsbruck, Technikerstr.\ 25, 6020 Innsbruck, Austria}
\begin{abstract}
We explore the structure of multipartite quantum systems which are entangled in multiple degrees of freedom. We find necessary and sufficient conditions for the characterization of tripartite systems and necessary conditions for any number of parties. Furthermore we develop a framework of multi-level witnesses for efficient discrimination and quantification of multidimensional entanglement that is applicable for an arbitrary number of systems and dimensions.
\end{abstract}


\maketitle

Entanglement plays a fundamental role in various fields of research. In quantum information processing, because it is heavily involved in quantum communication protocols \cite{nielsen} and at the heart of one of the most likely implementable models of a quantum computer \cite{qc}. But also in condensed matter systems it has become clear that there is a strong connection between phase transitions in complex systems and entanglement \cite{phase}.

The possible structure of correlations in large quantum systems is, however, so undeniably complex that little progress has been made on a general characterization of multipartite entanglement (for a good overview consult e.g. Refs.~\cite{multiq1,multiq2,multiq3}). Especially for systems that go beyond two degrees of freedom (and thus cannot be represented by qubits) little is known about the general structure of correlations.
In this paper we explore the involvement of different dimensions in multipartite entanglement. This is an interesting question from the theoretical point of view as this explains how many degrees of freedom (i.\ e.\ quantum levels) need to be effectively entangled to prepare a state and provides a measure of entanglement. Moreover, from a more practical point of view, entanglement among more than two levels allows to achieve further and/or more efficiently quantum information tasks \cite{expcrypto} and recent experiments are focusing on creating high dimensional entanglement \cite{dimexp0,dimexp1,dimexp2,dimexp3,dimexp4,dimexp5,dimexp6}. It is therefore desirable to put up easily testable conditions that can assure that systems of a certain entanglement dimensionality have been prepared and that the experimental data cannot be reproduced by entangling systems of lower dimensionality.

Whereas in the bipartite case a single number, the Schmidt number \cite{sn}, is sufficient to fully characterize the dimensionality of a given quantum state, the situation is more involved in the multipartite case as one needs to take into account multiple such numbers to characterize the state. To see this, consider pure states. In the bipartite case, the Schmidt rank, which is the rank of the reduced state of one particle (notice that they are both equal), answers the question of how many dimensions are necessary to faithfully represent the state and its correlations in any local basis. However, in a tripartite system there are three one-party reduced states that can essentially be of different rank and thus the question of how many degrees of freedom are involved has at least three answers. Determining which configurations of local ranks are possible can be regarded as a particular instance of the quantum marginal problem \cite{marginalp}, where it has been shown that local parameters provide information on global entanglement properties \cite{polytop}. Investigations on generalizations of the Schmidt rank are also interesting in the context of quantum computation. Although it has been shown that certain implementations of well-known algorithms require (multipartite) entanglement \cite{qa}, universal quantum computation is possible with little entanglement with respect to most bipartite measures \cite{maarten}. Interestingly, the Schmidt rank is an exception, being actually necessary for a quantum speed-up \cite{maarten,vidal}.

Recent papers \cite{multisn} have studied multipartite entanglement dimensionality under the assumption that all local ranks are equal, leading to a situation analogous to the bipartite case in which a single figure of merit is enough. We study in full detail for the first time the general case in this article, which is organized as follows. First, we present the natural generalization for the Schmidt rank and number for multipartite systems: the Schmidt rank and number vector. Then we show and illustrate how one can fully characterize the state space of tripartite quantum systems. Although more involved than in the bipartite case, it turns out that there is an underlying structure in the sets of states of different entanglement dimensionality. This allows us to introduce a general construction method for multi-level nonlinear witnesses, that can efficiently discriminate and quantify the entanglement dimensionality in arbitrary multipartite states.\\
Let us start by quickly reviewing the case of bipartite states composed of two subsystems $A$ and $B$. Let $\psi$ be a pure bipartite state and let $\rho_A$ and $\rho_B$ be its corresponding reductions. The Schmidt rank of $\psi$, is defined as $d_\psi:=\textrm{rank}(\rho_A)=\textrm{rank}(\rho_B)$. Since $d_\psi$ is the minimal number of terms one needs to write the state in a biorthogonal product basis (i.\ e.\ Schmidt decomposition), this number clearly gives the minimal local dimensions for subsystems $A$ and $B$. Thus, $\psi$ is effectively a two-qu$d$it state. The generalization to mixed states $\rho$ is given by the Schmidt number \cite{sn}
\begin{equation}
d_\rho= \min_{\mathcal{D}(\rho)}\max_{\{\psi_{i}\}}d_{\psi_{i}},
\end{equation}
where the minimization is over all ensemble decompositions of $\rho$, $\mathcal{D}(\rho)=\{p_i,|\psi_i\rangle : \rho=\sum_ip_i|\psi_i\rangle\langle\psi_i|\}$. This is a very natural definition as this means that $\rho$ cannot be obtained by mixing pure states of Schmidt rank lower than $d_\rho$ and that there exists a way to prepare the state by mixing states with Schmidt rank at most $d_\rho$. Moreover, the Schmidt number is an entanglement monotone and can thus be used to quantify the degree of entanglement \cite{sn} and, also, it can be operationally interpreted as the zero-error entanglement cost in the protocol of one-shot entanglement dilution \cite{BuscemiDatta}. Notice that although in general the computation of the Schmidt number is involved, there exist ways to obtain lower bounds for this measure. In particular, one can define the set of states with Schmidt number at most $d$, $S_d$, which induces a Russian doll structure of convex sets (i.\ e.\ $S_d\subset S_{d+1}$) and Schmidt number witnesses can be defined \cite{snwit}.

Let us now move to the multipartite case. For the sake of readability we will discuss the tripartite case in detail, as the generalization to even higher numbers of parties follows in a straightforward way. A tripartite pure state has three single-particle marginals of inequivalent rank (which we will from now on abbreviate via $\text{rank}(\rho_M):=r_M$), i.e. for $|\psi\rangle\in\mathcal{H}_A\otimes\mathcal{H}_B\otimes\mathcal{H}_C$ we can look at reduced states $\rho_i:=\text{Tr}_{\overline{i}}(|\psi\rangle\langle\psi|)$. Three out of the six possible reductions are sufficient as of course $r_A=r_{BC}$, $r_B=r_{AC}$ and $r_C=r_{AB}$ holds. Thus in order to characterize tripartite states three numbers are enough, i.e. $(r_A,r_B,r_C)$. Although these three ranks can potentially be different not every combination of integers can actually be achieved by a physical quantum state. One can show that the subadditivity of the R{\'e}nyi $0$-entropy, which translates as the submultiplicativity of the ranks is actually a necessary and sufficient constraint on the three numbers, i.e. let without loss of generality $r_A\geq r_B,r_C$ be fulfilled, then for every set of numbers fulfilling $r_A\leq r_B r_C$ there exists a pure state realizing exactly this combination \footnote{Necessity follows from $r_A=r_{BC}\leq r_Br_C$. For sufficiency it is enough to consider the state $\frac{1}{\sqrt{r_Br_C}}\sum_{m=0}^{r_B-1}\sum_{n=0}^{r_C-1}|m(r_C-1)+n\rangle|m\rangle|n\rangle$ which fulfills $r_A=r_Br_C$. Any state for which $r_A< r_Br_C$ can be constructed in this way by neglecting some terms in the sum.}. For a higher number of parties this is not sufficient anymore \footnote{A trivial counterexample would be $r_A=r_B=r_C=r_D=2$ and $r_{AB}=r_{AC}=r_{BC}=1$. It obeys the submultiplicativity of ranks, yet it is clearly impossible to realize.}, which could be solved via introducing the following conjecture for tripartite reductions of 4-partite pure states $r_{AB}r_{AC}r_{BC}\geq r_{A}r_{B}r_{C}$ \footnote{This inequality clearly holds if the tripartite state is pure with equality, for full rank states the left hand side is strictly greater and for all classical distributions it simply follows from the monotonicity of the $0$-entropy. It would also rule out the counterexample above.}.\\
Focusing again on tripartite systems we arrange $r_A$, $r_B$ and $r_C$ in non-increasing order to form the vector $(r_1,r_2,r_3)$. Given a pure state $|\psi\rangle$, its entanglement dimensionality vector (or Schmidt rank vector) is defined as $d_\psi:=(r_1^\psi,r_2^\psi,r_3^\psi)$. The extension of this definition to mixed states is not completely straightforward because, contrary to the bipartite case, if one consider states of entanglement dimensionality $r_1$ or less, $r_2$ or less and $r_3$ or less, this just defines a partial ordering and, as a consequence, one cannot trivially obtain a structure of sets in which, given any two subsets, one is always embedded in the other. This can be seen by considering the example of a state with Schmidt vector $(4,2,2)$ and a state of Schmidt vector $(3,3,2)$. In order to resolve this ambiguity, to obtain a well-defined mathematical structure and to impose a physically-meaningful classification we propose the following definition for entanglement dimensionality vectors (or Schmidt number vectors) for mixed states: A state $\rho$ has Schmidt number vector $d_\rho=(r_1,r_2,r_3)$ iff
\begin{equation}\label{snumbervec}
r_j=\min_{\mathcal{D}(\rho)}\max_{\{\psi_{i}\}}r_j^{\psi_{i}}.
\end{equation}
That is, for all ensemble decompositions of $\rho$, $\mathcal{D}(\rho)=\{p_i,|\psi_i\rangle : \rho=\sum_ip_i|\psi_i\rangle\langle\psi_i|\}$, there exists a $|\psi_i\rangle$ with $r_j^{\psi_i}$ at least $r_j$ and there exists a particular ensemble decomposition in which all $|\psi_i\rangle$ satisfy $r_j^{\psi_i}\leq r_j$ $\forall i$.

\begin{figure}[h!]
\begin{center}
  \includegraphics[scale=1.2]{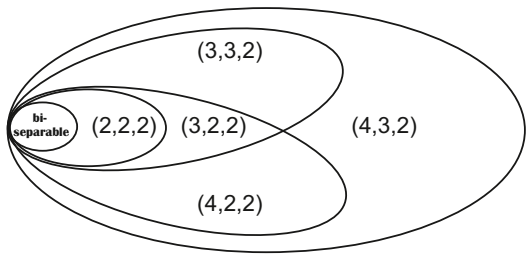}
  \caption{Schematic representation of a few sets of states with a given Schmidt number vector.}{\label{sets}}
\end{center}
\end{figure}

The structure of sets of states induced by this definition is depicted in Fig.\ \ref{sets}, where in a slight abuse of notation we denote by $(r_1,r_2,r_3)$ the set of all states with Schmidt number vector with entries at most $r_1$, at most $r_2$ and at most $r_3$. Some comments are in order. First, notice that each entry $r_j$ of the Schmidt number vector is an entanglement monotone. This is straightforward as the local rank of each $|\psi_i\rangle$ in an ensemble decomposition of $\rho$ cannot be increased by LOCC \cite{LoPopescu}. However, this is just a partial order as there exist incomparable states according to this measure, e.\ g.\ those in the subsets $(r_1,r_2,r_3)$ and $(r_1',r_2',r_3')$ when $r_1> r_1'$ but $r_2< r_2'$, which is very natural since the states in these subsets are LOCC incomparable. This is reflected in the structure of the set of states by the fact that there are subsets in which neither is included in the other like in the case of $(4,2,2)$ and $(3,3,2)$ as schematically shown in Fig.\ \ref{sets}. Second, it should be noticed that $d_\rho=(r_1,r_2,r_3)$ does not imply that $\rho$ has an optimal ensemble decomposition with one $|\psi_i\rangle$ such that $d_{\psi_i}=(r_1,r_2,r_3)$ but rather that the state cannot be written \textit{solely} as mixture of states which are all contained in a set which is lower in the hierarchy induced by the Schmidt number vector to $(r_1,r_2,r_3)$. Consider for instance the qudit ($d=7$) state $\rho=p|\psi_{332}\rangle\langle\psi_{332}|+(1-p)|\psi_{422}\rangle\langle\psi_{422}|$ with $0<p<1$ and
\begin{align}
|\psi_{332}\rangle&=\frac{1}{\sqrt{3}}(|000\rangle+|111\rangle+|122\rangle),\nonumber\\
|\psi_{422}\rangle&=\frac{1}{2}(|333\rangle+|344\rangle+|435\rangle+|446\rangle).\label{ex234}
\end{align}
This decomposition is clearly optimal as $|\psi_{332}\rangle\langle\psi_{332}|$ and $|\psi_{422}\rangle\langle\psi_{422}|$ are supported on orthogonal subspaces. Therefore, $d_\rho=(4,3,2)$ although it is a mixture of $(3,3,2)$ and $(4,2,2)$ states and does not contain any $(4,3,2)$ state in its support. However, this convention turns out to be very natural from the physical point of view when interpreting the Schmidt number vector as an indication of the number of levels one has to be able to effectively entangle to prepare the state. Despite it is not necessary to mix $(4,3,2)$ pure states to prepare $\rho$, this state cannot be obtained in an experiment without the ability to effectively access 4 quantum levels for one subsystem, 3 for another and 2 for the remaining one \footnote{One can also interpret this definition as the least Schmidt number vector possible if one considers only mixtures of comparable states.}. Third and last, one cannot exclude the possibility of a state that admits two different ensemble decompositions, each of which with states in incomparable subsets like $(3,3,2)$ and $(4,2,2)$. According to our definition such a state would have entanglement dimensionality $(3,2,2)$. This is again physically reasonable taking into account the entanglement monotonicity of each $r_j$. Moreover, this quantifies the least number of levels one must be able to effectively entangle.

\begin{figure}[th!]
\begin{center}
  \includegraphics[scale=0.6]{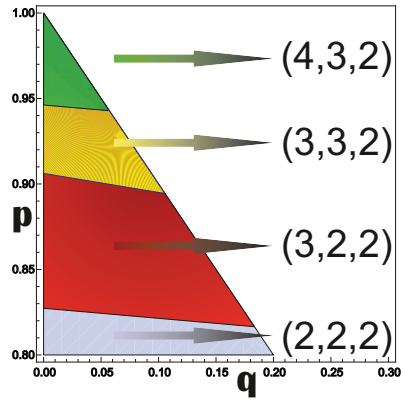}
  \caption{(Color online) Here we depict the noise resistance of our detection method in the presence of white and dephasing noise for the exemplary state from eq(\ref{examplestate}). In the top left corner the state is pure, going down vertically adds white noise and going diagonally to the right dephases the state. The differently colored(shaded) regions are labeled with the corresponding lower bounds on the dimensionalities that one can prove using the nonlinear witness from eq.(\ref{bounds}).}{\label{example1}}
\end{center}
\end{figure}

We consider now how to derive conditions to discriminate the entanglement dimensionality of a given mixed state of $N$ parties. Notice that, contrary to the bipartite case, although Schmidt number vector witnesses can be defined, a single one of them cannot fully identify certain states due to the lack of a Russian doll structure of convex sets. This can be seen by considering the states in $(4,3,2)$ which lie in the convex hull of $(3,3,2)$ and $(4,2,2)$. This problem can be overcome in principle by considering several entanglement witnesses or by defining nonlinear witnesses. We will follow this second approach by introducing the measures
\begin{equation}
E_k:=\inf_{\mathcal{D}(\rho)}\sum_ip_i S_k(\psi_i)\quad (k=1,2,\ldots,N).
\end{equation}
Here, $S_k(\psi)$ are the entries arranged in non-increasing order of the entropy vector given by the entropies of the single-particle reduced density matrices. For the sake of mathematical convenience we use the linear entropy, i.\ e.\ $S(\rho_A)=\sqrt{2(1-\Tr(\rho_A^2))}$. We will denote by $\rho_{s_1},\ldots,\rho_{s_N}$ the different single-party reduced density matrices in such a way that $S_k(\psi)=S(\rho_{s_k})$. The last component of this vector is equivalent to a measure of genuine multipartite entanglement that has been introduced in Ref.~\cite{Maetal} and intensively studied in Ref.~\cite{bound}. Notice that if $E_k>\sqrt{2(1-1/r)}$, this implies that $r_k\geq r+1$, i.e. we can lower bound $r_k\geq \lceil\frac{2}{2-E_k^2}\rceil$. Therefore, although the converse is not true, these measures can be used to obtain lower bounds for the Schmidt number vector, thus allowing the possibility of inferring that at least a certain entanglement dimensionality has been achieved. Actually, the measures $E_k$ are hard to compute in practice \cite{gurvits}; however, for simple measures and bipartite systems there exist techniques to estimate certain measures of entanglement using experimentally friendly witness techniques \cite{guhnewit,eisertwit}. In the following we derive a general framework that allows for the construction of nonlinear witnesses that are experimentally feasible and able to lower bound each $E_k$ and thus reveal even the non-convex structures of multipartite and multidimensional entanglement. In order to do that, let us first consider pure states, which we expand in the computational basis, $|\psi\rangle=\sum_\eta c_\eta|\eta\rangle$, with $\eta$ a multiindex of $N$ entries taking the values $0$ and $d-1$. It can be seen that
$S(\rho_{s_k})^2=\sum_{\eta,\eta'}|c_{\eta}c_{\eta'}-c_{\eta_{s_k}}c_{\eta_{s'_k}}|^2$, 
where the pair $(c_{\eta_{s_k}},c_{\eta_{s'_k}})$ is just equal to the pair $(c_{\eta},c_{\eta'})$, but with all components of $\eta$ and $\eta'$ that are part of the reduction $s_k$ exchanged. Using that $|C|\sum_{i\in C}|a_i|^2\geq|\sum_{i\in C}a_i|^2$ \cite{bound} and that $|a-b|\geq|a|-|b|$, we have that
\begin{equation}
S(\rho_{s_k})\geq \frac{1}{\sqrt{|C_k|}} \sum_{\eta,\eta'\in C_k}(|c_{\eta}c_{\eta'}|-|c_{\eta_{s_k}}c_{\eta_{s'_k}}|)
\end{equation}
for any subset $C_k$ of multiindices of $N$ entries. Therefore, we can bound our measures $E_k$ for pure states as
\begin{align}
E_k(\psi)\geq \frac{1}{\sqrt{C_k}} \sum_{\eta,\eta'\in C_k} (|c_{\eta}c_{\eta'}|-\min_{\{s_m\}}\sum_{m=1}^k|c_{\eta_{s_m}}c_{\eta_{s_m}'}|).
\end{align}
Now we can extend this to mixed states via the observation that $\inf(A-B)\geq\inf A-\sup B$. First, it is clear that
\begin{align}
\inf_{\mathcal{D(\rho)}}\sum_i p_i|c^i_{\eta}c^i_{\eta'}|\geq|\sum_i p_ic^i_{\eta}{c^i_{\eta'}}^*|=|\langle\eta|\rho|\eta'\rangle|.
\end{align}
For the supremum we can use
\begin{align}
\sup_{\mathcal{D(\rho)}}&\sum_ip_i\min_{\{s_m\}}\sum_{m=1}^k|c^i_{\eta_{s_m}}c^i_{\eta_{s_m}'}|\nonumber\\
&\leq\min_{\{s_m\}}\sup_{\mathcal{D(\rho)}}\sum_ip_i\sum_{m=1}^k|c^i_{\eta_{s_m}}c^i_{\eta_{s_m}'}|\nonumber\\
&\leq \min_{\{s_m\}} \sum_{m=1}^k\sqrt{(\sum_ip_i|c^i_{\eta_{s_m}}|^2)(\sum_ip_i|c^i_{\eta_{s_m}'}|^2)}\nonumber\\
&=\min_{\{s_m\}} \sum_{m=1}^k\sqrt{\langle\eta_{s_m}|\rho|\eta_{s_m}\rangle\langle\eta'_{s_m}|\rho|\eta'_{s_m}\rangle}.
\end{align}
In conclusion, we end up with $E_k(\rho)\geq W_k(\rho)$, where
\begin{align}
 W_k(\rho)&:=\frac{1}{\sqrt{|C_k|}}\sum_{\eta,\eta'\in C_k}\left[|\langle\eta|\rho|\eta'\rangle|\right.\nonumber\\
 &\left.-\min_{\{s_m\}} \sum_{m=1}^k\sqrt{\langle\eta_{s_m}|\rho|\eta_{s_m}\rangle\langle\eta'_{s_m}|\rho|\eta'_{s_m}\rangle}\right].\label{bounds}
\end{align}
Thus, we obtain easily computable lower bounds on the Schmidt number vector in terms of the entries of the density matrix. Notice that we are free to play with the subsets $C_k$ of entries to be considered to obtain the most stringent bounds. Also, the conditions are basis-dependent and one can furthermore optimize over all possible choices of local bases. \\
It is crucial to investigate how these lower bound nonlinear witness vectors enable a dimensionality classification in the presence of noise. Typically one encounters either white noise or dephasing in experimental situations. So let us consider the following state
\begin{align}
\rho_{test}=p\rho_{(4,3,2)}+q\rho_{dp}+\frac{1-p-q}{64}\mathbbm{1}\,,\label{examplestate}
\end{align}
where $\rho_{432}=|\psi_{432}\rangle\langle\psi_{432}|$ is our multidimensionally multipartite entangled target state
\begin{align}
|\psi_{432}\rangle=\frac{1}{2}(|000\rangle+|111\rangle+|012\rangle+|123\rangle)\,,
\end{align}
and $\rho_{dp}$ is the completely dephased state. The crucial step in using the nonlinear witness element as a lower bound on the entropy, and thus the dimensionality is of course the selection of the sets $(\eta,\eta')\in C_k$. We now use a different choice for each entry of the witness vector in order to achieve good noise resistance. For the first component we choose $C_1=\{(000,111),(000,123),(012,123)\}$, for the second we choose $C_2=\{(000,111),(000,123),(012,123),(000,012),(111,123)\}$ and for the maximum entropy we can use the full set $C_3=\{(000,111),(000,123),(012,123),$ $(000,012),(111,123),(111,012)\}$. Then using Eq.\ (\ref{bounds}) we arrive at an analytical expression for the entropy lower bounds which we plot in Fig.{\ref{example1}}\\
What is also clearly visible in the example is the fact that using the linear entropy lower bounds to determine the dimensionality of course works best if the distribution of the eigenvalues of the marginals is rather flat. Although this might present itself as a weakness if one aims to characterize fully the dimensionality of mixed states on a theoretical level, we would like to argue that this method actually is advantageous for all practical purposes. First we want to point out that just as in the bipartite case there exist a lot of full rank, indeed even arbitrary dimensionality, states that are $\epsilon$-close to the separable states, so even if such a state were to be detected by a more precise criterion, it would immediately introduce problems with experimental precision that would make a meaningful distinction impossible. Secondly the entanglement entropy is at the heart of the advantage of higher dimensional systems, e.g. it directly determines the size of the generated key in a bipartite quantum key distribution scenario (see e.g. Refs.~\cite{expcrypto,badrnd}). Using our lower bounds one can achieve two things, first to give a reliable detection method for the dimensionality of multipartite systems and at the same time answer how useful these extra dimensions are in terms of potential applications.\\
In conclusion we have presented for the first time a general classification of multipartite entanglement in terms of multidimensional entanglement. We give necessary and sufficient conditions for the existence of tripartite entanglement classes and necessary conditions for any number of parties and with this illustrate the structure of multipartite and multidimensional entanglement and the partial hierarchy of subsets of states it induces. Furthermore we develop a framework of entropy-vector lower bounds that employ nonlinear witness techniques. We explicitly show that these techniques work very well in exeprimentally feasible and plausible scenarios.\\
We believe that this not only presents testable conditions about general quantum correlations that are the heart of quantum physics, but also may directly serve as security tests in multidimensional applications of entanglement in quantum key distribution systems.\\
Open challenges include the characterization of $n$-qudit state spaces and the relation of multidimensionality of entanglement with its distillability (maybe as an extension of the conjecture in Ref.~\cite{snwit}).
\noindent\emph{Acknowledgements.}
M.H. would like to acknowledge productive discussions and valuable input from Dagmar Bru{\ss}, Ottfried G{\"u}hne, Hermann Kampermann, Milan Mosonyi, Marcin Pawlowski, Marco Piani, Andreas Winter and Junyi Wu and especially the initial discussions with Matthias Kleinmann. Furthermore MH acknowledges funding from the EC-project IP "Q-Essence" , the ERC Advanced
Grant "IRQUAT" and the MC grant "Quacocos". J.I. de V. acknowledges financial support from
the Austrian Science Fund (FWF): Y535-N16 and F40-FoQus F4011-N16.

\end{document}